\documentclass[conference]{IEEEtran}

\usepackage{cite}
\usepackage{amsmath,amssymb,amsfonts}
\usepackage{algorithmic}
\usepackage{graphicx}
\usepackage{textcomp}
\usepackage{xcolor}
\usepackage{subfig}

\usepackage{soul}
\usepackage{comment}
\usepackage{float}
\usepackage[draft]{hyperref}

\IEEEoverridecommandlockouts    
\begin{document}


\title{Experimental Impact Analysis of Cyberattacks in Power Systems using Digital Real-Time Testbeds}
\author{
\IEEEauthorblockN{\textbf{Kalinath Katuri}\IEEEauthorrefmark{1}, \textbf{Ioannis Zografopoulos}\IEEEauthorrefmark{2},  \textbf{Ha Thi Nguyen}\IEEEauthorrefmark{1}, \textbf{Charalambos Konstantinou}\IEEEauthorrefmark{2}}
\IEEEauthorblockA{\IEEEauthorrefmark{1}Eversource Energy Center, Department of Electrical and Computer Engineering, University of Connecticut, Storrs, CT, USA\\
\IEEEauthorrefmark{2}CEMSE Division, King Abdullah University of Science and Technology (KAUST)
}
\IEEEauthorblockA{
E-mail: 
\{kalinath.katuri, ha.t.nguyen\}@uconn.edu, 
\{ioannis.zografopoulos, charalambos.konstantinou\}@kaust.edu.sa }
}
          
\maketitle
\vspace{-10cm}
\begin{abstract}
Smart grid advancements and the increased integration of digital devices have transformed the existing power grid into a cyber-physical energy system. This reshaping of the current power system can make it vulnerable to cyberattacks, which could cause irreversible damage to the energy infrastructure resulting in the loss of power, equipment damage, etc. Constant threats emphasize the importance of cybersecurity investigations. At the same time, developing cyber-physical system (CPS) simulation testbeds is crucial for vulnerability assessment and the implementation and validation of security solutions. In this paper, two separate real-time CPS testbeds are developed based on the availability of local research facilities for impact analysis of denial-of-service (DoS) attacks on microgrids. The two configurations are implemented using two different digital real-time simulator systems, one using the real-time digital simulator (RTDS) with a hardware-in-the-loop (HIL) setup and the other one using OPAL-RT with ExataCPS to emulate the cyber-layer infrastructure. Both testbeds demonstrate the impact of DoS attacks on microgrid control and protection operation.
\end{abstract}


\section{Introduction}
The integration of computing and communication resources in power systems, with the goal of developing a robust smart grid infrastructure, is transforming power transmission, distribution, and utilization. The quest for secure and resilient energy power systems (EPS) has promoted the deployment of distributed and decentralized grid architectures. According to the 2020 U.S. Department of Energy smart grid system report \cite{paper1}, smart grids are the application of digital and cyber infrastructure to the physical system in order to perform the functions of sensing, communications, control, computing, and data and information management to assist planning and operations. Reliable communication is required to leverage the benefits of smart grids and the support services provided by distributed energy resources (DER). As a result, the cyber layer of EPS and information and communication technologies (ICT) are becoming an integral part of the smart grid paradigm.

As the power grid transitions to a cyber-physical system, it becomes highly dependent on digital communication systems, increasing the risk of cyberattacks. Some of the most recent examples of cyberattacks on the power grid include the one in Ukraine on December 23, 2015, which resulted in the loss of power to more than 225,000 customers and affected more than 50 substations \cite{paper2}, and the attack on the nuclear power plant in India during 2019 \cite{paper3}. 
The greater threat to cyber-physical energy systems (CPES) and the prohibitive cost and operational risks of conducting cybersecurity investigations on actual power systems motivate the need for CPES testbeds \cite{zografopoulos2022distributed}.
Such testbed setups can capture the spatial and temporal characteristics of distributed EPS, while co-simulation studies can comprehensively evaluate the impacts of cyber contingencies (e.g., communication delays) on the physical system's dynamic performance \cite{zografopoulos2021cyber}.

Research teams around the world have developed diverse CPES testbed topologies to mirror real-world applications and enable cybersecurity research \cite{cintuglu2016survey, 7428063}. In \cite{paper5}, a real-time cyber-physical testbed with Real Time Digital Simulator (RTDS) and OPNET is developed to investigate the impact of man-in-the-middle (MITM) attacks on the 11-Bus system. In \cite{paper4}, a testbed 
was developed with both real-time and non-real-time capabilities. The authors use RTDS and PowerFactory for the physical system components, i.e.,  RTDS is communicated to the cyber layer through IEC 61850 and DNP3, and PowerFactory is connected to a virtual substation through an OPC interface. Researchers at Sandia National Laboratories have developed a virtual control system testbed that incorporates various commercially available components for the cyberattack impact analysis on the energy critical infrastructure \cite{paper6}. A comprehensive survey of industrial control system testbeds and their capabilities, along with the essential data required to perform cybersecurity studies, is furnished in \cite{conti2021survey}. \looseness=-1

The security investigations performed on cyber-physical testbeds include, \textit{i)} vulnerability research, \textit{ii)} mitigation research, \textit{iii)} security validation, and \textit{iv)} impact analysis \cite{paper4, zografopoulos2021security}. In \cite{paper7}, the impact of coordinated powerbot attacks on the power system is studied, while \cite{paper8} and \cite{paper9} demonstrate the impact of denial-of-service (DoS) attacks on power grids. The impact of cross-layer firmware attacks on microgrid (MG) inverter controllers is discussed in \cite{zografopoulos2022time}, along with methodologies to design host-based defenses to overcome such incidents. In \cite{paper10}, cyberattacks against the IEC 61850 protocol (in photovoltaic (PV) installations) are studied. In \cite{jahromi2019cyber} and \cite{10065529}, the authors present the impacts of communication-based attacks targeting CPES protection systems. The results of the aforementioned research underline the importance of CPES security studies. Equally important is the design and constant improvement of unified cyber-physical simulation and/or hardware platforms where we can investigate different threats, varying system topologies, and operation conditions. 

This paper presents two different real-time simulation platforms for CPES to evaluate the impact of diverse cyberattack scenarios on EPS. The work illustrates the investigation of cybersecurity studies on different platforms based on the availability of local research facilities. Specifically, our contributions are as follows:
\begin{itemize} 
\item Impact analysis of cyberattacks targeting distribution systems is conducted with one of the testbeds developed by University of Connecticut (UCONN)'s Eversource energy center using RTDS Novacor chasis and a hardware relay.
\item The consequences of communication delays -- due to DoS conditions -- are illustrated using toolkits provided by \emph{OPAL-RT} and \emph{ExataCPS} in King Abdullah University of Science and Technology (KAUST)'s CPES testbed. 
\item Development of a hardware-in-the-loop (HIL) implementation to investigate the overcurrent conditions of a PV system and a battery under the impacts of malicious cyberattacks degrading critical operations.
\end{itemize}

The rest of the paper is organized as follows. Section \ref{s:experimental_setup} provides an overview of the testbed setups at UCONN and KAUST and their capabilities. The impact of trip delays and setpoint changes is demonstrated via HIL simulations in Section \ref{s:RTDS}. The impact of attacks targeting the grid's cyber infrastructure and their effects on generation and frequency are presented in Section \ref{s:Opal_cases}. Finally, Section \ref{s:conclusion} concludes the paper by highlighting the importance of CPES testbeds for future secure power systems.
\section{Experimental Laboratory setup} \label{s:experimental_setup}
This section outlines the implementation of two different experimental laboratory setups for the impact analysis of MGs under cyberattacks. These testbeds are built leveraging real-time simulation platforms.

\subsection{HIL Testbed with RTDS at UCONN}
\begin{figure}[t]
\centering
    \begin{tabular}{ c @{\hspace{6pt}} c }
    \includegraphics[width=0.49\linewidth, height=0.4\linewidth]{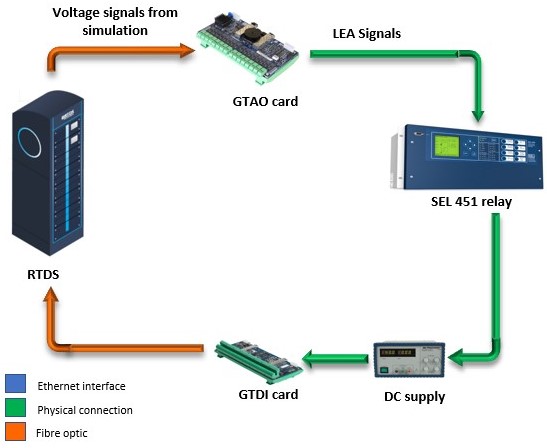} &
    \includegraphics[width=0.49\linewidth, height=0.4\linewidth]{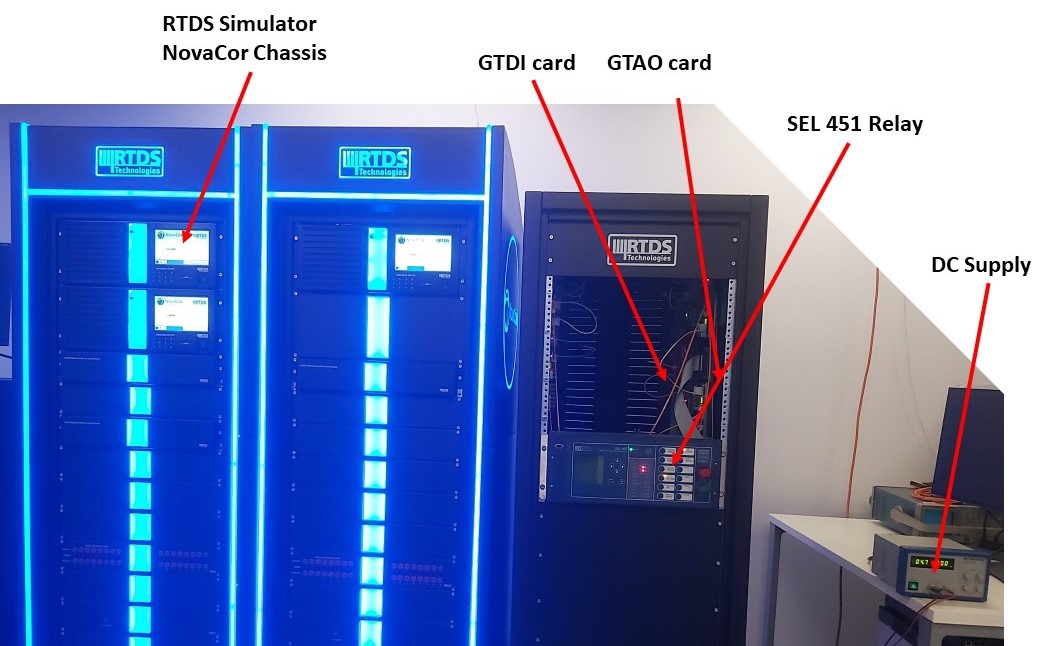} \\
    \small a) Schematic &
    \small b) Lab setup
    \end{tabular}
    \caption{Hardware-in-the-loop (HIL) testbed with RTDS.}
    \label{fig:HiL}
\end{figure}

{The CPES testbed is developed on the RTDS platform. 
For the impact analysis, a controller HIL (CHIL)-based testbed is preferred not only to evaluate the logic or settings of a protection system but also to assess closed-loop controller performance. The schematic and lab setup of HIL developed for the relay under test are shown in Fig. \ref{fig:HiL}}.

The SEL 451 relay hardware in this work is used as a non-directional over-current relay. The analog output card (GTAO) of the RTDS is used to retrieve the voltage and current signals from the simulation. And these signals are connected directly to the relay at a peak-to-peak voltage of 10 V. The relay output is connected to the digital input (GTDI) card of RTDS (through a 5V DC supply). Two cyberattack scenarios -- DoS and MITM attacks -- are simulated, and their impacts on the MG and the relay performance are evaluated in real-time.
\subsection{Co-simulation Testbed with OPAL-RT at KAUST}
\begin{figure}[t]
    \centering\includegraphics[width=0.8\linewidth]{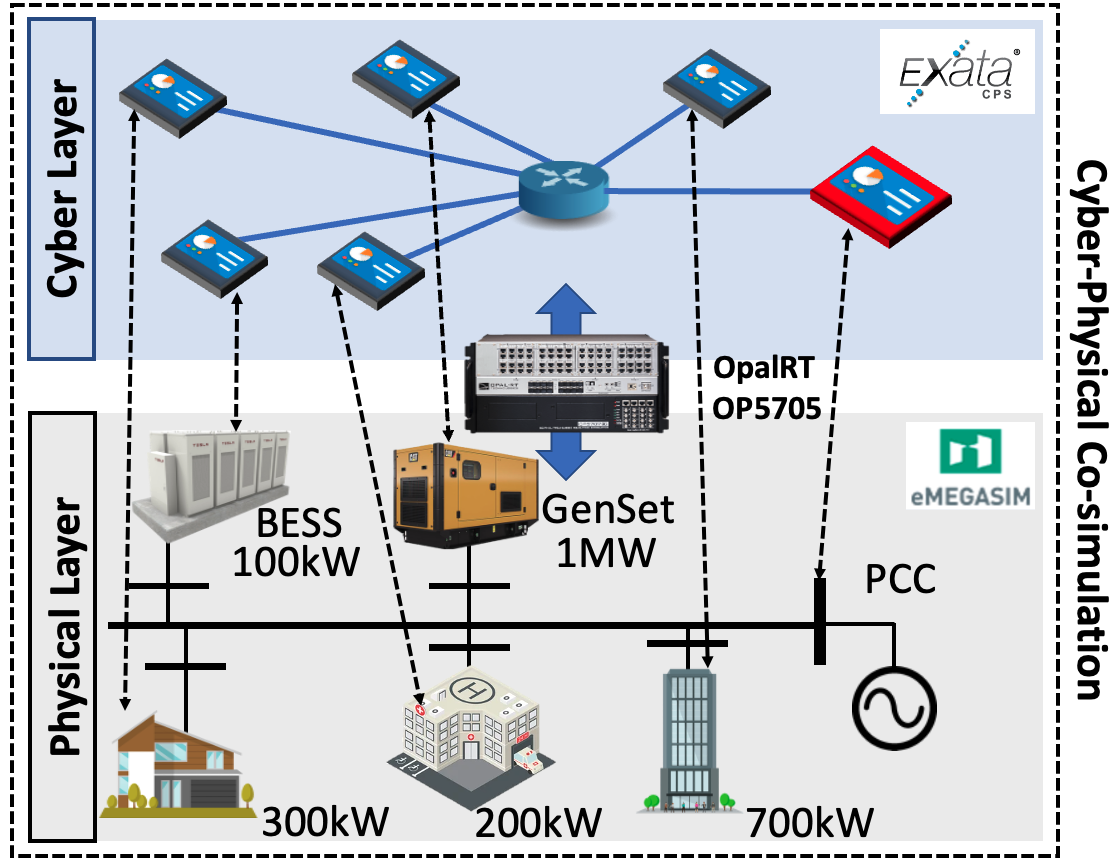}
    \caption{Cyber-physical co-simulation setup with OPAL-RT.}
    \label{fig:phys_system}
    \vspace{-2mm}
\end{figure}

This real-time co-simulation testbed is composed of two distinct layers, as shown in Fig. \ref{fig:phys_system}. The setup investigates the impact that different durations of DoS attacks could have on the physical portion of the EPS.

\subsubsection{Physical Model} \label{s:physical}
On the physical layer, we model a MG in which we distinguish two power sources; namely, a diesel generator (Genset) and a battery energy storage system (BESS) with nameplate capacities of 1 MW and 100 kW, respectively. Apart from the generation buses, we have also included three load buses, a non-critical 300 kW lumped residential load, a 200 kW critical load (e.g., hospital), and a commercial load with a power demand of 700 kW. Both generation and load components are interfaced through the point-of-common-coupling (PCC), where the MG central controller is also located, with the main grid. The aforementioned physical layer is simulated using the \emph{OPAL-RT} real-time simulator and the \emph{eMegasim} power electronics toolkit.

\subsubsection{Cyber Model} \label{s:cyber}
On the cyber layer, for each of the buses (generation or load) residing on the physical layer, a corresponding cyber layer node is emulated. We use the IEEE-1815 Distributed Network Protocol 3 (\emph{DNP3}) for the communication between the cyber layer nodes. Furthermore, it can be seen from Fig. \ref{fig:phys_system}, that the cyber layer nodes are connected in a radial architecture where the MG central controller at the PCC serves as the \emph{DNP3} master, and the remaining nodes constitute the \emph{DNP3} outstations. For the modeling and simulation of the cyber layer, we leverage the \emph{EXataCPS} toolkit that allows the emulation of communication networks and can be seamlessly integrated within the \emph{OPAL-RT} environment.

\begin{figure}[t]
    \centering
    \includegraphics[width=\linewidth]{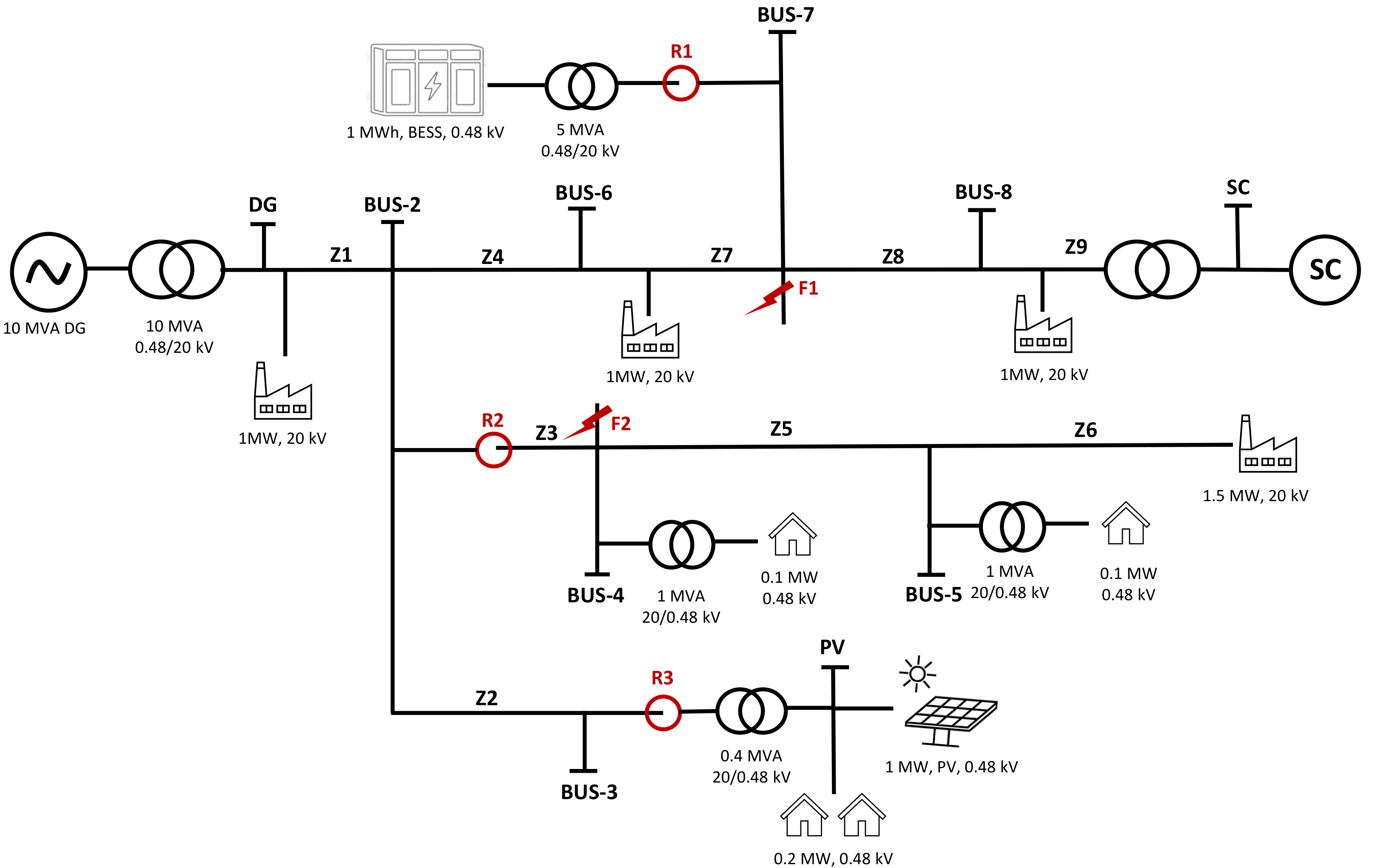}
    \caption{Single line diagram of the microgrid network simulated in RTDS.}
    \vspace{-0.1in}
    \label{fig:Microgrid}
\end{figure}

\section{RTDS-HIL Testbed: Cases \& Results} \label{s:RTDS}
The MG simulated in this work consists of a 20 kV distribution network with two types of DERs, an aggregated representation of rooftop PV of 1 MW, and a large-scale BESS of 1MW capacity. The single-line diagram (SLD) of the MG implemented in the RTDS platform is shown in Fig. \ref{fig:Microgrid}. Through a 10 MVA, 20/0.4 kV transformer, a 0.48 kV Genset is connected to the network. A total load of 4 MW is connected on both the 20 kV and 0.48 kV levels. In this work, the attacker is considered to have gained access to the communication network, and the test cases are the possible scenarios of DoS and MITM attacks on the network.

The impacts of cyberattacks are studied by imitating the attack's physical behavior on the RTDS network. Three different test cases (under attack) are considered. The first two cases imitate the behavior of a DoS attack during a three-phase short-circuit fault at F1 and F2 locations, and the third case imitates the behavior of a MITM attack on the PV system. The physical impact of all three cases is studied with the help of the SEL 451 relay used in the CHIL experiment, connected at locations R1, R2, and R3, respectively \cite{paper11}. Relays are connected through a current transformer as shown in Fig. \ref{fig:Microgrid}. For these test cases, three-phase short-circuit faults are simulated due to their severe fault currents.

\subsection{Battery Trip Delay Impact on Microgrid}
Through this test case, the impact of DoS attacks on a protection trip during a three-phase short-circuit fault at location F1 on Bus-7 is studied. During the fault, due to the low short-circuit fault path, the battery will contribute to the fault current, resulting in the tripping of the relay connected to the high voltage (HV) side of the transformer at the R1 location, which is referred to as ``sympathetic tripping''.
\begin{figure}[t]
    \centering
    \includegraphics[width=0.95\linewidth]{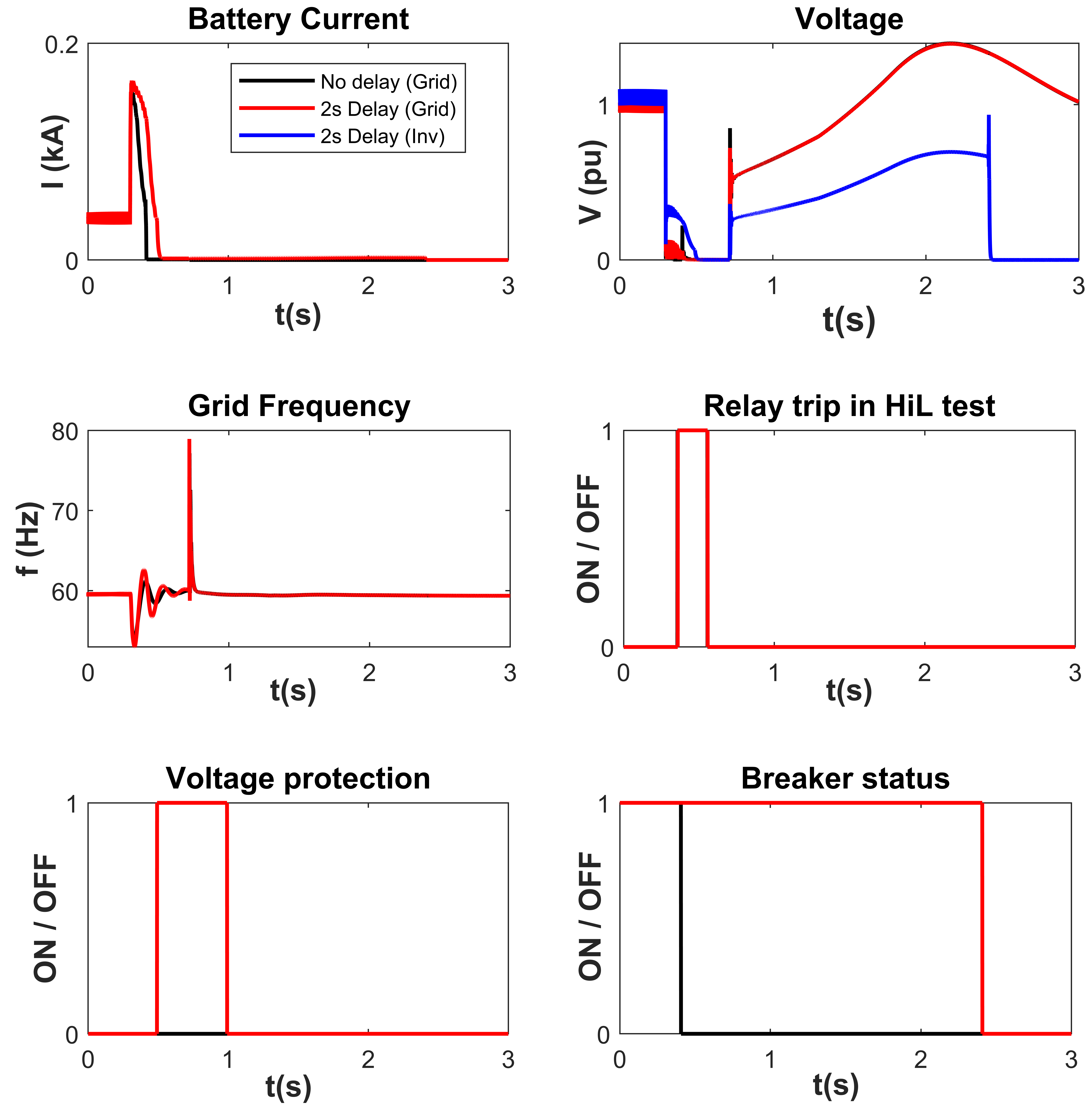}
    \caption{Battery test case.}
    \vspace{-0.1in}
    \label{fig:BESS}
\end{figure}
To study the impact of a DoS attack during this fault, it is assumed that the trip command from the relay to the breaker is delayed, which causes a delay in isolating the battery from the faulted network. The impact of this delayed breaker operation is simulated by interfering with the relay output from SEL 451 in the CHIL simulation. The results are plotted in Fig. \ref{fig:BESS}. A $2$ $secs$ delay is assumed in the breaker operation.


The comparison of the relay trip is shown in Fig. \ref{fig:BESS} with normal operation denoted using black and under cyberattack in red. The current contribution from the BESS during the fault is momentary, as the fault is temporary in nature. It is observed that, without any delay, the voltage at PCC is recovered at a slower rate after the fault is cleared. However, this does not trigger any voltage protection, and the system regains stability after the fault. In the $2$ $secs$ delay case, even though the frequency stabilizes to its nominal value after a few cycles, the inverter voltage fails to recover to the rated 1.0 pu range, resulting in the activation of the voltage protection on the BESS side. 

\subsection{Delay in Relay Operation for Feeder Faults}
This test case is used to examine the effects of operations being delayed as a result of a DoS attack during a fault in the middle feeder. The main overcurrent feeder protection relay is connected at R2 and operates in this capacity. It is expected that a coordinated DoS attack can cause the trip feedback to the relay at R2 to be delayed. The delayed response of the relay, and consequently the circuit breaker under attack, can cause disturbances in the voltage and frequencies of the diesel generator, hence affecting other connected equipment in the MG. If the fault persists, this delay in relay operation would result in unrecoverable damage to the system, depending on its duration. The prolonged unstable voltage and frequency profiles may also result in the tripping of inverter-based resources connected to the system, which could potentially lead to MG collapse.

\begin{figure}[t]
    \centering
    \includegraphics[width=0.95\linewidth]{Figures/Feeder5.png}
    \caption{Genset response to the delay in fault on feeder case.}
    \label{fig:Feeder}
    \vspace{-3mm}
\end{figure}

\begin{figure}[t]
    \centering
    \includegraphics[width=0.95\linewidth]{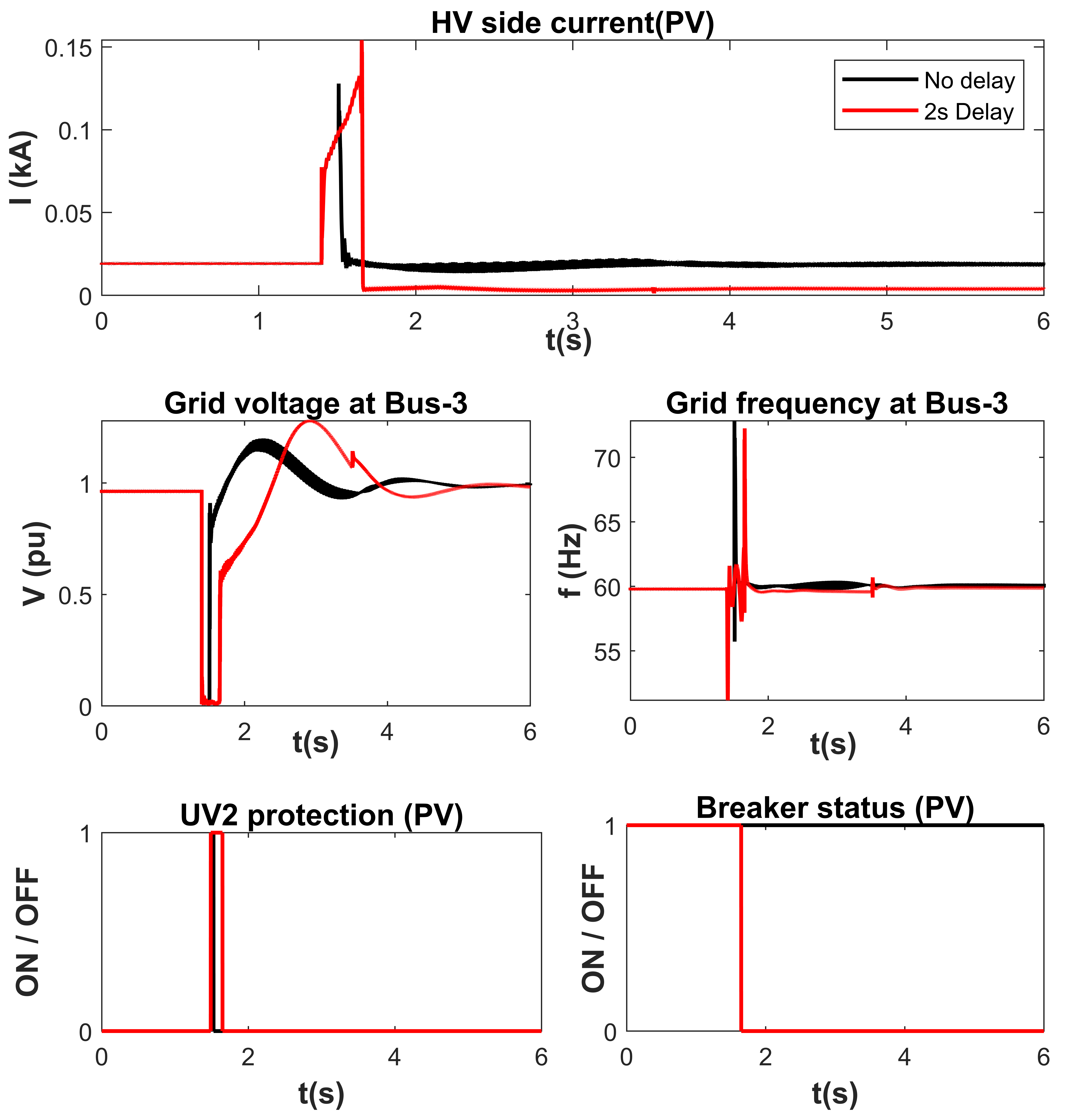}
    \caption{PV response to fault on middle feeder.}
    \label{fig:PVB4}
\end{figure}
\begin{figure}[t]
    \centering
    \includegraphics[width=0.95\linewidth]{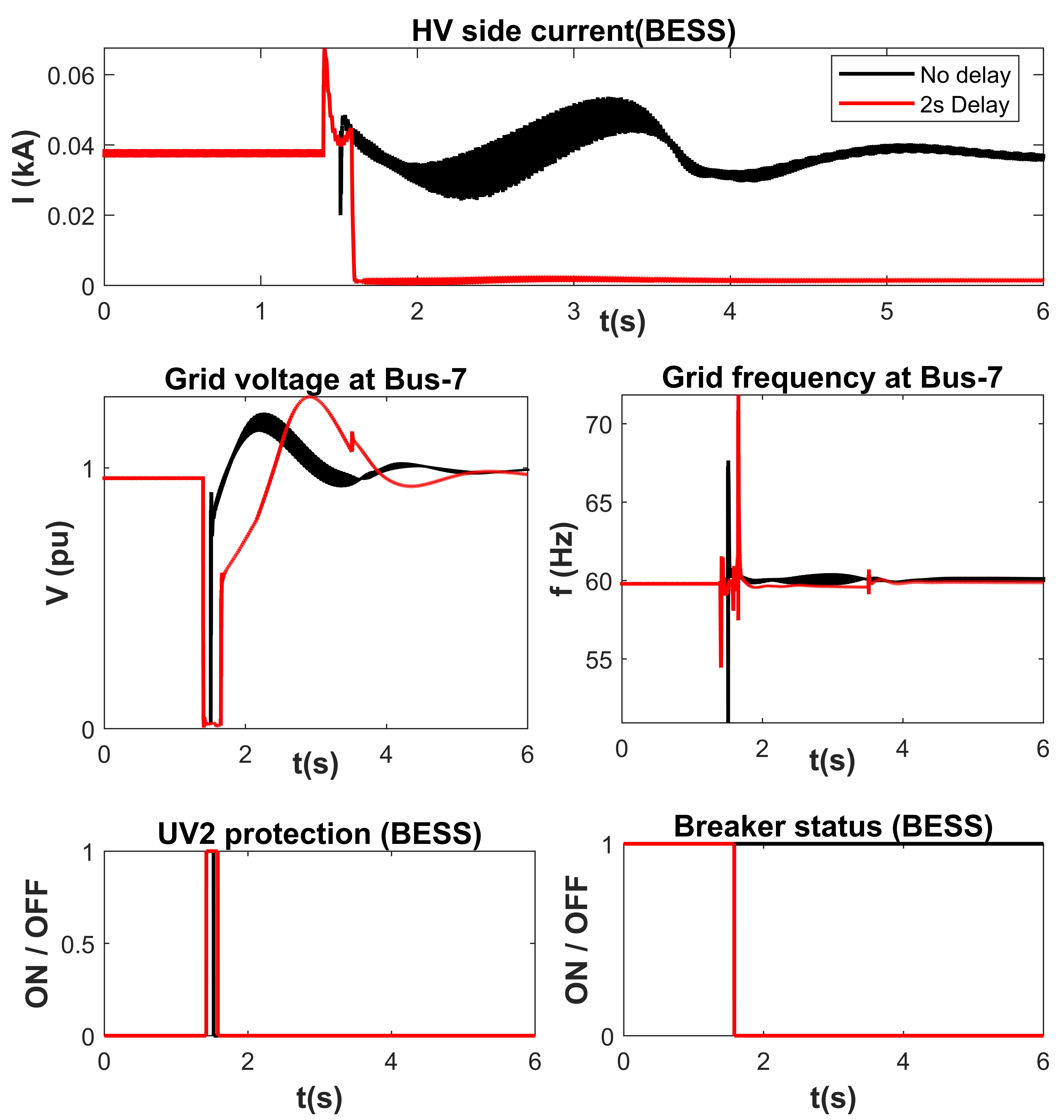}
    \caption{BESS response to fault on middle feeder.}
    \vspace{-0.1in}
    \label{fig:BESSB4}
\end{figure}

Fig. \ref{fig:Feeder} shows the results for the $2$ $secs$ delayed impact (red) compared to the no delay (black) response for the fault in Bus-4 at F2. The red curves show that the frequency of the Genset exhibits a large oscillatory behavior, and the system voltage on Bus-2 is observed to be abnormal for a considerable period. These results not only show the severe nature of the impact on the Genset, but also give an understanding of the damage to voltage-sensitive equipment connected to the network. Fig. \ref{fig:PVB4} depicts the operation of voltage protection for the PV-connected bus. Fig. \ref{fig:BESSB4} shows the impact of delayed breaker operation in the system. \textcolor{black}{Both the BESS and the PV systems have tripped on voltage protection as a result of the voltages recorded at Bus-7 and Bus-3 falling below the set value of voltage protection for UV2 ($<$ 0.45 pu) for longer than the set period ($0.16$ $secs$). The voltage values used are compliant with the recommendations provided in IEEE 1547:2018.}


\subsection{PQ Setpoint Change for PV System}
In the secondary control of the PV plant, active (P) and reactive (Q) power is controlled by the utility set points. Such set points enable the effective use of green energy available at any time of the day. Through this case study, the impact of MITM attacks on PV systems is studied. When an attacker gains access to the control network, they can either deny the service or change the control set points of the PV plant without the knowledge of the utility. This deliberate action would result in a sudden increase in reverse power flow to the distribution grid. This sudden change in reference setpoints of P and Q would result in an increase in the current flow from the PV to the network, which can trigger the overcurrent relay connected at R3 at the PCC point for the PV plant in Fig. \ref{fig:Microgrid}. Furthermore, it may also activate internal voltage and frequency protections depending on the system state during the change. 
\begin{figure}[t]
    \centering
    \includegraphics[width=0.95\linewidth]{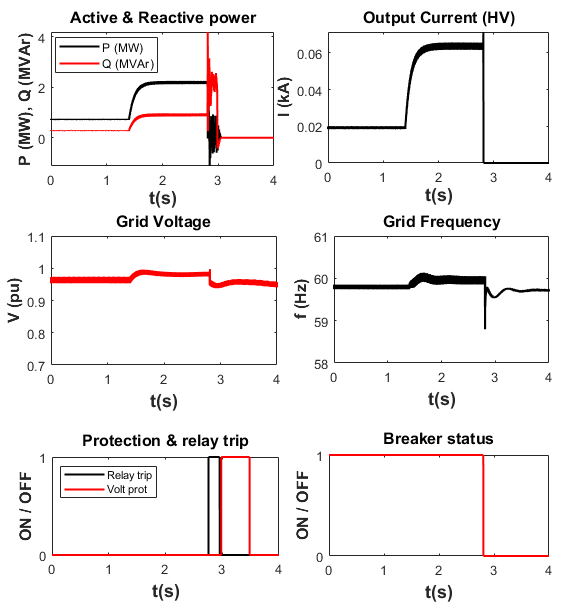}
    \caption{PV response to PQ set point change.}
    \vspace{-1mm}
    \label{fig:PVcase}
\end{figure}

The results in Fig. \ref{fig:PVcase} show the system conditions when the P and Q outputs of the PV system increase. The effect is demonstrated not only by the higher current output but also depicted in the voltage profile. The current change activates the overcurrent relay, subsequently resulting in breaker operation, whereas the voltage profile results in voltage protection pickup. The spikes in P and Q profiles are related to the disconnection of the PV system during the breaker operation.

\section{OPAL-RT Testbed: Cases \& Results} \label{s:Opal_cases}
\vspace{-1mm}
In the following case study, we leverage a real-time co-simulation testbed to examine the cross-layer impact of DoS cyberattacks. Specifically, high-fidelity models are considered for both the cyber layer (accounting for communication delays) and the physical layers. Then, using the real-time co-simulation environment (Fig. \ref{fig:phys_system}), we demonstrate how the impact of different DoS conditions (encountered on the cyber layer) can propagate to the actual physical grid compromising its stable operation and affecting its steady-state stability.

We evaluate different DoS attack examples and their impact on the EPS frequency and power measurements. DoS cyberattacks, impeding the timely delivery of critical control signals (e.g., islanding or load-shedding commands), can introduce abnormal system states and lead to severe system maloperation. In our simulations, the DoS severity is indicated by the duration of the introduced time-delay attacks, which can, in turn, have minor (e.g., uneconomic operation if power flow data are delayed) or major (e.g., generation instabilities if control commands do not respect the constraints of the EPS) impacts on system objectives. It should be noted that although the discussed attacks (Section \ref{s:DoS_results}) are performed solely on the cyber layer (by delaying communications), their impact on grid properties (frequency, generated power, etc.) can be severe. \textcolor{black}{In the rest of the section, we present simulation results illustrating the impact of DoS attacks on the isolated MG setup. The specifics about the MG's physical and cyber modeling are detailed in Sections \ref{s:physical} and \ref{s:cyber}.} 

\begin{figure}[t]
\centering
    \subfloat[]{
        \includegraphics[width=0.8\linewidth, height=0.5\linewidth]{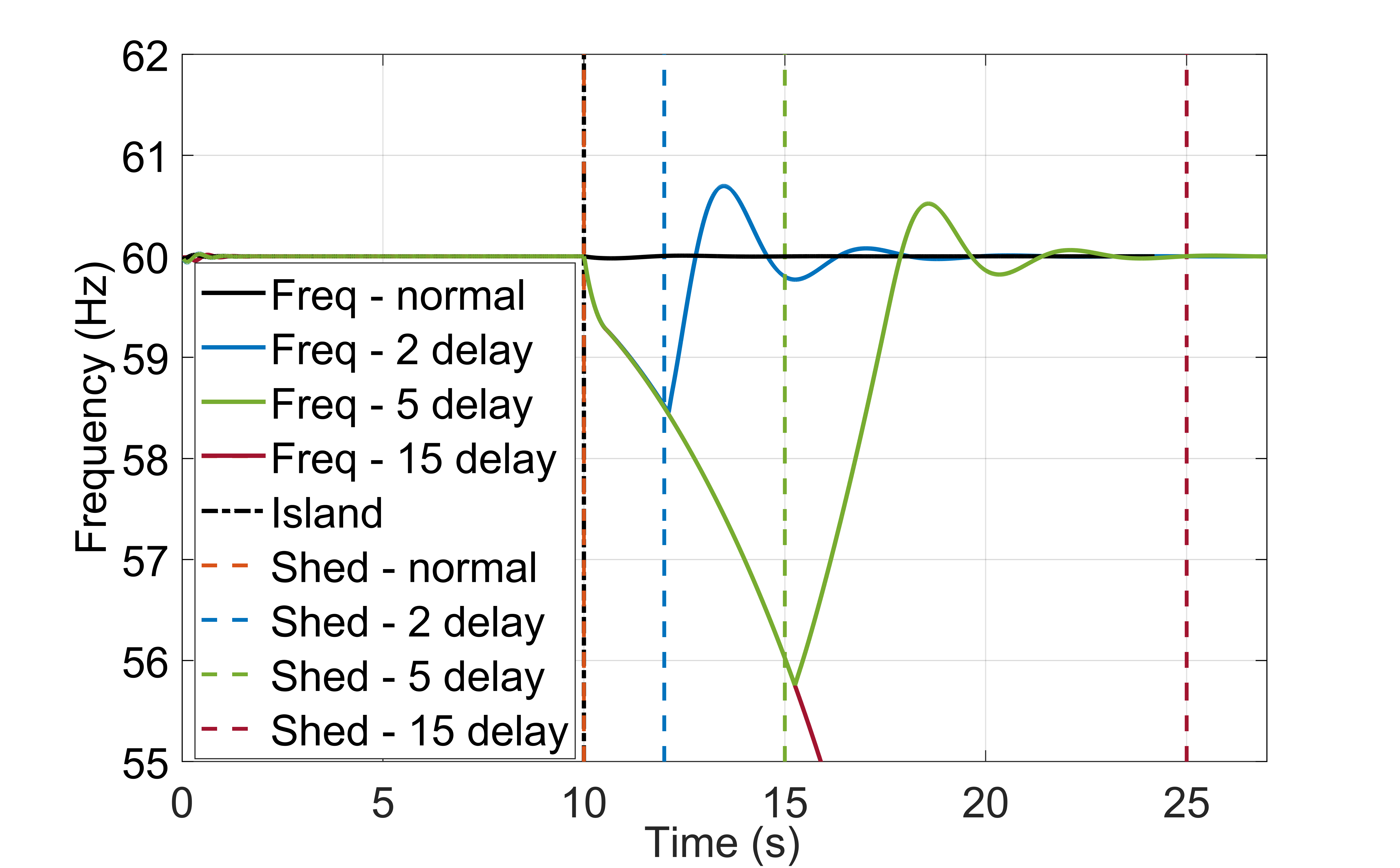}
        \label{fig:freq_all}
    } \\
    \subfloat[]{
        \includegraphics[width=0.8\linewidth, height=0.5\linewidth]{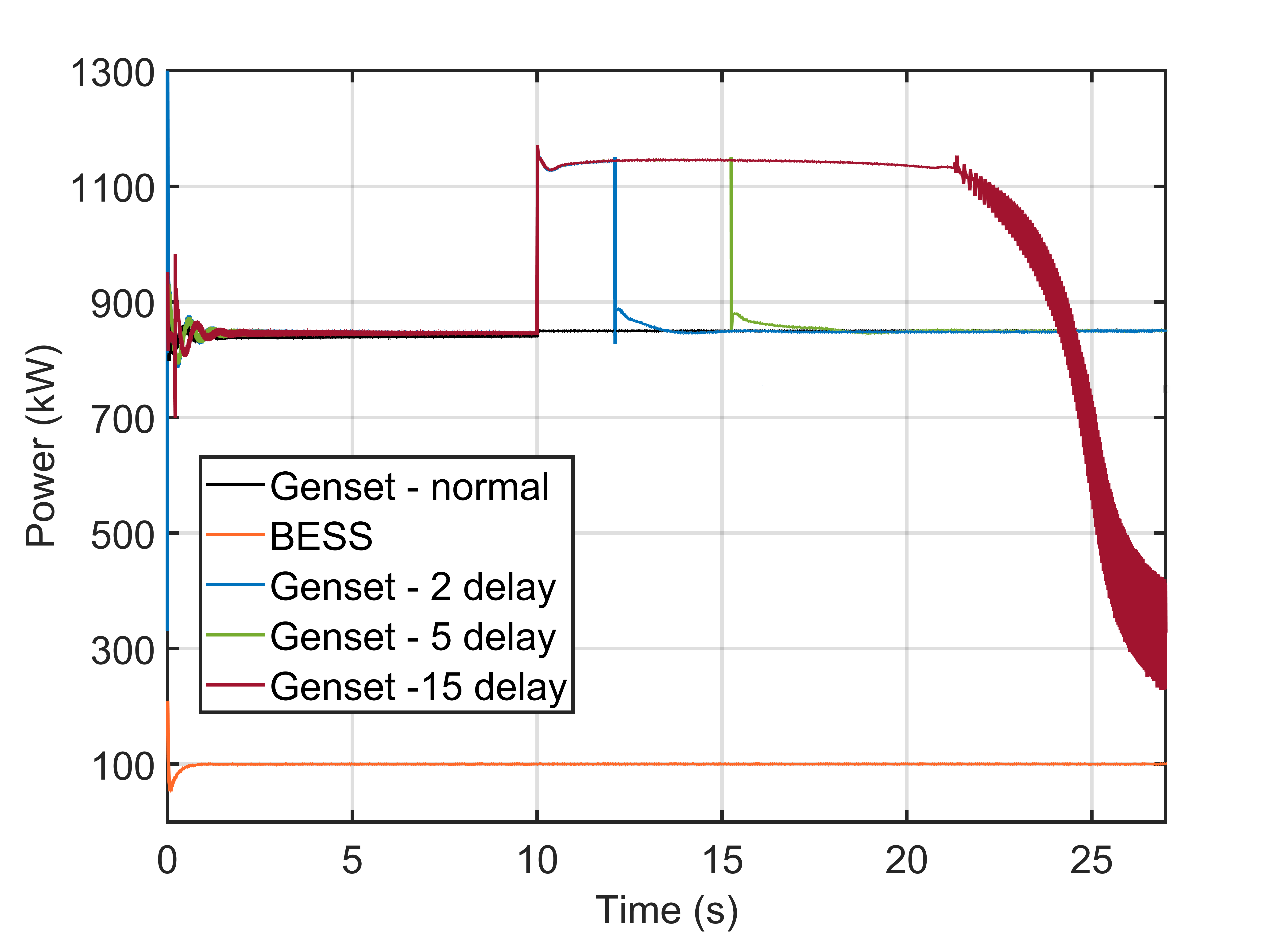}
        \label{fig:pow_all}
    } \\
   
\caption[CR]{Frequency and power measurements of the MG during DoS attacks. 
\subref{fig:freq_all} Impact of $2$, $5$, and $15$ $secs$ load-shedding delay on frequency, \subref{fig:pow_all}  
Impact 
on the power generation profiles. 
\looseness=-1} 
\label{fig:freqPow_measurements}
\vspace{-1mm}
\end{figure}
\subsection{Simulation Results} \label{s:DoS_results}

In this part, we present the potential impact of cyber-layer-based attacks (introducing communication delays) on MG operations. Fig. \ref{fig:freqPow_measurements} illustrates the frequency fluctuations and power sharing between the generator and BESS during three different DoS attack scenarios. In more detail, the MG system (Fig. \ref{fig:phys_system}) is islanded from the main grid by tripping the circuit breaker located at the PCC at $t=10$ $secs$. The inherent MG generation capacity, i.e., power sourced from the Genset and the BESS, cannot meet the aggregated load demand, as can happen in weak MG architectures. To avoid under-frequency conditions, defensive load-shedding should be performed, prioritizing critical and commercial loads (as described in Section \ref{s:physical}) and curtailing the power demand introduced by residential loads. 

Immediately after the MG isolation ($t=10$ $secs$), the MG controller issues a load-shedding command to maintain grid frequency stability. However, at the same time, an adversary performs a DoS attack that ``floods'' the network's communication capacity. As a result, the DNP3 messages issued by the MG master controller are delivered with arbitrary delays to the outstation devices (e.g., load controllers). In our case, the load-shedding command for the residential load is delayed. Fig. \ref{fig:freq_all} illustrates the impact of such time DoS delays on the MG frequency if the load-shedding mechanism is delayed by $2, 5,$ and $15~secs$, respectively. 

Even in the less severe DoS attack case, where the load-shedding command is delayed for $2~secs$, we can observe that the MG frequency drops below the nominal operational range, e.g., $59.5$~Hz \cite{nercfreq}. In the second scenario, where the load-shedding command is delayed by $5~secs$, a significant frequency decrease occurs too, i.e., the frequency drops below $56$~Hz, in which case corrective mechanisms should be immediately enforced to protect the grid assets from this abnormal frequency condition. It should be noted that regardless of the insufficient MG power generation, in both of the aforementioned scenarios, the MG is able to maintain its stable operation, although the Genset has to momentarily operate above its rated capacity for the duration of the DoS. 

The MG resilience to DoS-based cyberattacks is compromised in the last scenario. Fig. \ref{fig:freq_all} and Fig. \ref{fig:pow_all} illustrate the impact on frequency and generator stability during an attack that delays the load-shedding command delivery for $15~secs$. \textcolor{black}{The results in both graphs explicitly demonstrate the potential impacts that cyber contingencies, e.g., communication delays, could impose on the MG dynamics. In more detail, in Fig. \ref{fig:freq_all}, we illustrate that the MG frequency drops below $55$~Hz if the command is delayed by more than $5~secs$, which could prompt further load-shedding to maintain frequency stability.} Furthermore, as shown in Fig. \ref{fig:pow_all}, the Genset is operated above its nominal capacity for more than $10~secs$, eventually rendering the whole MG system unstable. At this point, the MG system -- currently operating autonomously -- would experience a blackout and would not be able to support the $200kW$ critical load. 
\vspace{-2mm}
\section{Conclusion} \label{s:conclusion}
\vspace{-1mm}
The risk of cyberattacks on power grids and energy infrastructure has been continuously increasing. The advancement of smart grid infrastructure poses a greater risk of cyberattacks due to unknown vulnerabilities. The growing importance of cybersecurity research has motivated researchers to develop unique testbeds focused on different aspects of cybersecurity research. \looseness=-1

In this work, two different testbeds are studied using real-time simulations for the impact analysis of cyberattacks on MGs. The two testbeds show that cybersecurity research can be conducted on different platforms based on the availability of local facilities. The first setup implements a HIL experiment using the RTDS simulator, while the second focuses on a cyber-physical setup using the \emph{OPAL-RT} simulator. Both configurations investigate DoS attack impacts in real-time environments, and the results from both testbeds show that cyberattacks can cause damage and even cascading failures or power system collapse. More research on CPES security is essential to provide insights for system operators and propose solutions to prevent, detect, and mitigate impacts. In future work, different types of cyberattacks, such as MITM and replay attacks, will be studied on both testbeds to evaluate the impacts of these attacks on power grids and provide potential mitigation strategies.

\bibliographystyle{IEEEtran}
\bibliography{refs}
\end{document}